\documentclass[twocolumn,showkeys,pre]{revtex4}
\usepackage{bm}% bold math

\begin{document}

\title{The ${\cal VW}$ transformation--A Simple Alternative to the Wilson GF Method}

\author{George D. J. Phillies}
\affiliation{Department of Physics, WPI, Worcester MA 01609}

\begin{abstract}
An alternative, the ${\cal VW}$ transformation, is proposed to replace the Wilson GF method for calculating molecular vibration frequencies and normal modes. The method yields precisely the same eigenmodes and and eigenfrequencies that are found with the GF method. The transformation proceeds entirely in the mass-normalized Cartesian coordinates of the individual atoms, with no transformations to internal coordinates.  The ${\cal VW}$ transformation thus offers an enormous computational simplification over the GF method, namely the mathematical apparatus needed to transform to internal coordinates is eliminated.  All need for new researchers to understand the complex matrix transformations underlying the GF method is thus also removed. The ${\cal VW}$ transformation is not a projection method; the internal vibrations remain dispersed over all $3N$ atomic coordinates.  In the ${\cal VW}$  method, the $3N \times 3N$ force constant matrix ${\cal V}$ is replaced with a new $3N \times 3N$ matrix ${\cal W}$. The replacement is physically transparent.  The matrix ${\cal W}$ has the same internal normal modes and vibration frequencies that ${\cal V}$ does.  However, unlike  ${\cal V}$, ${\cal W}$ is not singular, so the normal modes and normal mode frequencies can be obtained using conventional matrix techniques.
\end{abstract}

%\pacs{31.15.-p,33.20.Ea,36.20.Ng,33.20.Tp}
\keywords{molecular vibrations, Wilson GF method}
\maketitle

\section{Introduction}

It has long been known that one may, to good approximation, calculate the vibrational spectrum of a molecule by means of classical mechanics.  In this calculation, the
molecule is treated as a set of coupled harmonic oscillators, the atoms being the oscillating masses and the chemical bonds servings as springs. Couplings are primarily provided by the covalent bonds linking the atoms together.  Displacing atoms from their equilibrium positions changes bond lengths and angles, thereby creating harmonic restoring forces.  The orthodox eigenvalue-eigenvector description of coupled harmonic oscillators then yields molecular vibration frequencies and vibrational modes.

For an $N$-atom molecule, there are $3N$ atomic coordinates and therefore $3N$ potential normal modes. A mathematical complication arises in the search for these modes.  A molecule in a gas or liquid is free to translate or rotate without changing the relative positions of its atoms.  Translating or rotating an isolated molecule therefore does not create a restoring force that would oppose the translation or rotation. The matrix inversion approach to finding the eigenvalues and eigenvectors that describe the molecular vibrations immediately encounters a fatal difficulty. The relevant matrix is singular. Some of its eigenvalues are zero. The conventional solution process is inapplicable.

A half-century ago, Wilson overcame this difficulty by introducing the GF method\cite{wilsonGF}.  In his approach, the internal vibrations of a molecule are projected from the original $3N$-dimension space into a $3N-6$-dimensional subspace. Excluded from the subspace are six dimensions corresponding the whole-molecule translations and rotations, these being the dimensions that create the singular matrix challenge.  The $(3N-6) \times (3N-6)$ matrix corresponding to the projected subspace is non-singular and can be solved for the eigenvalues and eigenvectors corresponding to the molecule's internal vibrations. Historically, the Wilson GF method offered an additional practical advantage.  With the limited computational facilities of a half-century ago, the labor difference in inverting a $3N-6$ rather than a $3N$ dimensional matrix was non-trivial.

This note proposes an alternative to the Wilson GF method.  The method yields precisely the same eigenmodes and and eigenfrequencies that are found with the GF method. The new method retains real-space normalized Cartesian atomic coordinates at all steps in the calculation.  The corresponding price -- one must invert a $3N \times 3N$ rather than a $(3N-6)\times (3N-6)$ dimensional matrix -- has due to computational advances over the past half-century largely ceased to be of significance.  Section II of this paper treats notational preliminaries.  Section III presents the actual method.  A Discussion appears as Section IV.

\section{Notational Preliminaries}

This section presents the needed notation. The molecule has $N$ atoms $(1, 2, 3, \ldots, N)$ and total mass $M$.  Atom $i$ has mass $M_{i}$ and position $\mathbf{R}_{i}$.
The molecular center-of-mass location is  $\mathbf{R}_{\textrm cm} = \sum_{i} M_{i} \mathbf{R}_{i} /M$.  The atomic coordinates with respect to the molecular center of
mass are
\begin{equation}
      \mathbf{s}_{i} = \mathbf{R}_{i} - \mathbf{R}_{\textrm cm}
      \label{eq:localcoordinates}
\end{equation}
The molecular moment-of-inertia tensor is
\begin{equation}
      \mathcal{M} = \sum_{i=1}^{N} M_{i} \mathbf{s}_{i} \otimes  \mathbf{s}_{i}
\end{equation}
with $\otimes$ denoting the vector outer product. The three eigenvectors of $\mathcal{M}$ are are the molecule's principal axes $\mathbf{\Omega}_{j}$ for $j = 1, 2, 3$; the unit vectors corresponding to the $\mathbf{\Omega}_{j}$ are the $\hat{\mathbf{\Omega}}_{j}$.

Displacements of the atoms from their equilibrium positions are represented by $N$ vectors $\mathbf{X}_{i} \equiv (X_{i}, Y_{i}, Z_{i})$.  More compressed expressions are sometimes obtained by writing the displacements as a single $3N$-dimensional vector $\mathbf{x} \equiv {x_{j}}$, where $j \in (1, 3N)$. In particular, the potential
energy $U$ may be Taylor-series expanded around the equilibrium positions of the atoms as
\begin{equation}
      U(\mathbf{X}) = U^{(0)} + \sum_{j=1}^{3N} \frac{\partial U^{(0)}}{\partial x_{j}} x_{j} + \frac{1}{2}  \sum_{j,k=1}^{3N} \frac{\partial^{2} U^{(0)}}{\partial x_{j} \partial
      x_{k}} x_{j} x_{k} + \ldots.
      \label{eq:potentialenergymatrix}
\end{equation}
Here the superscript "0" refers to the evaluation of the potential energy, and its derivatives, at the equilibrium atomic positions.  At these positions, the first derivative terms all vanish.  The matrix of second derivatives is the force constant matrix, which is usefully denoted
\begin{equation}
      U_{jk} = \frac{\partial^{2} U^{(0)}}{ \partial x_{j}\partial x_{k}}.
      \label{eq:Ujkdef}
\end{equation}
The force $F_{j}$ corresponding to atomic coordinate $j$ is $- \partial U(\mathbf{X})/\partial x_{j}$.  To lowest non-zero order in the displacements, Newton's second
law can be written in terms of displacement coordinates as
\begin{equation}
        m_{j} \frac{\partial^{2} x_{j}}{\partial t^{2}} =  - \sum_{k=1}^{3N}  U_{jk} x_{k}.
        \label{eq:NewtonssecondX}
\end{equation}
Here $m_{j}$ is the mass corresponding to coordinate $j$ so that, e.g., $m_{1}=m_{2}=m_{3} \equiv M_{1}$.

To transform eq.\ \ref{eq:NewtonssecondX} to an eigenvalue-eigenvector equation, one introduces new mass-normalized coordinates $y_{j} = \sqrt{m_{j}} x_{j}$, with $\mathbf{y} = (y_{1},
y_{2}, \ldots y_{3N} )$, and a new symmetric matrix $\mathcal{V}$ whose components are
\begin{equation}
     V_{jk} = \frac{U_{jk}}{ (m_{j}m_{k})^{1/2}}.
\end{equation}
Rewriting eq.\ \ref{eq:NewtonssecondX} in terms of the new coordinates,
\begin{equation}
    \frac{\partial^{2} y_{j}}{\partial t^{2}} =  - \sum_{k=1}^{3N}  V_{jk} y_{k}.
     \label{eq:symmetricform}
\end{equation}

Included in the solutions to equation \ref{eq:symmetricform} are $3N-6$ normal modes corresponding to the internal vibrations.  The corresponding $3N-6$ normal mode frequencies are denoted $\omega_{n}$, the eigenvalues being the $-\omega_{n}^{2}$.  The $n^{\rm th}$ normal mode vector as represented in $\mathbf{y}$ coordinates is  $\mathbf{e}^{(n)}$; its $3N$ components are the  $e^{(n)}_{j}$. The
$\mathbf{e}^{(n)}$ are taken to be normalized, so that
\begin{equation}
        \sum_{j=1}^{3N} (e^{(n)}_{j})^{2} = 1.
        \label{eq:normalized}
\end{equation}
Taking the displacements in eq.\ \ref{eq:symmetricform} to correspond to a normal mode $\mathbf{e}^{(n)}$, one has
\begin{equation}
     - \omega_{n}^{2} e^{(n)}_{j} =  - \sum_{k=1}^{3N}  V_{jk} e^{(n)}_{k}.
     \label{eq:symmetricform2}
\end{equation}

The difficulty with extracting the $\mathbf{e}^{(n)}$ from $\mathcal{V}$ is that the matrix $\mathcal{V}$ is singular.  In addition to the $3N-6$ normal modes
corresponding to internal vibrations, $\mathcal{V}$ has six eigenvectors (three translations, three rotations) $\mathbf{f}^{(n)}$, whose eigenvalues are zero. The components of the $\mathbf{f}^{(n)}$ in $\mathbf{y}$ coordinates are the
$f^{(n)}_{k}$, for which
\begin{equation}
    0 =  - \sum_{k=1}^{3N}  V_{jk} f^{(n)}_{k}.
     \label{eq:symmetricform3}
\end{equation}
Because the $3N$ $\mathbf{e}^{(n)}$ and $\mathbf{f}^{(n)}$ are all orthogonal to each other, the change of coordinates used in the Wilson GF method can separate the $\mathbf{e}^{(n)}$ and $\mathbf{f}^{(n)}$ into two non-communicating subspaces.

\section{The Real-Space Solution Method}

Having considered the needed notation, the real-space solution method is now presented. The procedure is to replace the $3N \times 3N$ matrix $\mathcal{V}$ with a new $3N \times 3N$ matrix $\mathcal{W}$ matrix which (i) still has the $\mathbf{e}^{(n)}$ as $3N-6$ of its eigenvectors, (ii) which has the same $\omega_{n}$ as the respective eigenvalues of the $\mathbf{e}^{(n)}$, but (iii) \emph{which is not singular}.  Because $\mathcal{W}$ is not singular, it can be solved with standard matrix methods to determine the $\mathbf{e}^{(n)}$ and the $\omega_{n}$.  The replacement is possible because we know the $\mathbf{f}^{(n)}$ in advance.

The starting point of the real space solution method is that a real symmetric matrix can be written in terms of its eigenvalues and eigenvectors.  In particular, because
the $\mathbf{f}^{(n)}$ all have eigenvalue zero, the potential energy matrix $\mathcal{V}$ is correctly expressed
\begin{equation}
        \mathcal{V}  =  - \sum_{n=1}^{3N-6}  \omega_{n}^{2} \mathbf{e}^{(n)}  \otimes \mathbf{e}^{(n)}.
       \label{eq:matrixVform}
\end{equation}
Direct calculation invoking the orthogonality of the $\mathbf{e}^{(n)}$ confirms $\mathcal{V} \cdot \mathbf{e}^{(n)} =  - \omega_{n}^{2} \mathbf{e}^{(n)}$, as required. The individual matrices $\mathbf{e}^{(n)}  \otimes \mathbf{e}^{(n)}$ serve as filters, selecting out from an arbitrary set of displacements the extent to which each of the $3N$ modes and rigid-body displacements is present.

We do not yet know either the $\mathbf{e}^{(n)}$ or the $\omega_{n}$.  However, we do know what the six $\mathbf{f}^{(n)}$ are, namely  on purely physical grounds the $\mathbf{f}^{(n)}$ are the rigid-body translations and rotations of the molecule.  The $\mathbf{f}^{(n)}$ are used to construct from $\mathcal{V}$ a new matrix $\mathcal{W}$
\begin{equation}
        \mathcal{W}  = \mathcal{V}  + \sum_{n=1}^{6}  a_{n} \mathbf{f}^{(n)}  \otimes \mathbf{f}^{(n)}.
       \label{eq:matrixW1form}
\end{equation}
which replaces $\mathcal{V}$, namely
\begin{equation}
    \frac{\partial^{2} y_{j}}{\partial t^{2}} =  - \sum_{k=1}^{3N}  W_{jk} y_{k}.
     \label{eq:symmetricformA}
\end{equation}

The $a_{n}$ are six arbitrary unequal non-zero constants to be supplied by the user.  The $a_{n}$ are the six eigenvalues of $\mathcal{W}$ corresponding to the six eigenvectors $\mathbf{f}^{(n)}$. Because the $\mathbf{e}^{(n)}$ and the $\mathbf{f}^{(n)}$ are all orthogonal to each other, the $\mathbf{e}^{(n)}$ are seen by explicit calculation to be eigenvectors of $\mathcal{W}$.  By explicit construction and eqs.\ \ref{eq:matrixVform} and \ref{eq:matrixW1form}, the eigenvalues of $\mathcal{W}$ corresponding to the $\mathbf{e}^{(n)}$ are the correct $\omega_{n}^{2}$, unperturbed by the difference between $\mathcal{V}$ and $\mathcal{W}$. Finally, by direct construction $\mathcal{W}$ is a symmetric matrix with $3N$ non-zero eigenvalues, so $\mathcal{W}$ is non-singular, allowing solution of eq.\ \ref{eq:symmetricformA}  by standard methods.

The implicit $\mathcal{V} \cdot \mathbf{f}^{(n)} = 0$ is the physical requirement that the internal forces of a body cannot change the body's linear or angular momentum.  The $a_{n}$ are rationally chosen to support the numerical stability of the matrix inversion that now leads to the $\omega_{n}$ and the $\mathbf{e}^{(n)}$ without encountering any singularities, namely none of the $a_{n}$ should be equal to any of the $\omega_{n}^{2}$, and none of
the $a_{n}$ should be too far outside the range of values spanned by the  $\omega_{n}^{2}$.

To use the method, actual values are needed for the six $\mathbf{f}^{(n)}$.  The three translations are displacements on any three perpendicular axes, e.g., for the
$x$-translation as represented in $\mathbf{Y}$ coordinates, up to a normalization that has no effect on the method,
\begin{equation}
   \mathbf{f}^{(1)} = (\sqrt{M_{1}}, 0, 0, \sqrt{M_{2}}, 0, 0,\ldots,\sqrt{M_{N}},0,0),
\end{equation}
and similarly for $\mathbf{f}^{(2)}$ and $\mathbf{f}^{(3)}$. There is of course no physical requirement that the three translation eigenvectors be aligned with the three
Cartesian coordinate axes; any three orthogonal axes are equally acceptable. Eigenvectors parallel to the Cartesian axes are simply the easiest to generate. The rotation
eigenvectors are constructed from the $N$ vectors $\mathbf{s}_{i}$ and the three $\hat{\mathbf{\Omega}}_{j}$ as
\begin{equation}
     \mathbf{f}^{(3+j)} = \hat{\mathbf{\Omega}}_{j} \times (\sqrt{M_{1}} \mathbf{s}_{1}, \sqrt{M_{2}} \mathbf{s}_{2},\ldots,\sqrt{M_{N}} \mathbf{s}_{N}),
\end{equation}
the object in parentheses on the rhs being the $3N$-vector formed by concatenating the $N$ three-vectors $\sqrt{M_{j}}\mathbf{s}_{j}$.

\section{Discussion}

The objective here was to demonstrate an alternative to the Wilson GF method for calculating molecular vibration frequencies.  The
alternative is shown in the previous section.  One begins with $\mathcal{V}$, the mass-normalized matrix of the second derivatives of the potential energy surface.  The molecular vibrational modes and their frequencies are eigenvectors and eigenvalues of $\mathcal{V}$, but these eigenvectors and eigenvalues are inaccessible because $\mathcal{V}$ is singular: Six of its eigenvalues are zero.  Because the eigenvectors corresponding to the six vanishing eigenvalues are known and trivially calculated explicitly, a simple alternative to the Wilson GF method arises.  The matrix $\mathcal{V}$ and the six singular eigenvectors $\mathbf{f}^{(n)}$ are used to generate a new matrix $\mathcal{W}$ that: (i) has the same eigenvectors as $\mathcal{V}$, (ii) yields the same molecular vibration frequencies as  $\mathcal{V}$, but (iii) is non-singular, so it can be solved directly.

The mathematical method seen here is more generally useful, namely it can be applied to any similar problem in which one wants to calculate the eigenvectors and eigenvalues of a symmetric matrix, the matrix is singular, but the eigenvectors having eigenvalue zero are known explicitly.

\end{document}